\documentclass[aps,prl,twocolumn,superscriptaddress]{revtex4}

\usepackage{graphicx}
\usepackage{dcolumn}
\usepackage{bm}
\usepackage{amsmath}
\usepackage{amssymb}
\usepackage{latexsym}
\usepackage{epsfig}
\usepackage{amsbsy}
\usepackage{array}
\usepackage{amssymb}
\usepackage{bm}

\begin{document}

	\preprint{Physical Review Letters}
	
	\title{Realization of Superadiabatic Two-qubit Gates Using Parametric Modulation in Superconducting Circuits}

	\author{Ji Chu}
	\thanks{These authors contributed equally to this work.}
	\affiliation{National Laboratory of Solid State Microstructures, School of Physics,
		Nanjing University, Nanjing 210093, China}
	\author{Danyu Li}
	\thanks{These authors contributed equally to this work.}
	\affiliation{National Laboratory of Solid State Microstructures, School of Physics,
		Nanjing University, Nanjing 210093, China}
	\author{Xiaopei Yang}
	\thanks{These authors contributed equally to this work.}
	\affiliation{National Laboratory of Solid State Microstructures, School of Physics,
		Nanjing University, Nanjing 210093, China}
	
	\author{Shuqing Song}
	\affiliation{National Laboratory of Solid State Microstructures, School of Physics,
		Nanjing University, Nanjing 210093, China}
	\author{Zhikun Han}
	\affiliation{National Laboratory of Solid State Microstructures, School of Physics,
		Nanjing University, Nanjing 210093, China}
	\author{Zhen Yang}
	\affiliation{National Laboratory of Solid State Microstructures, School of Physics,
		Nanjing University, Nanjing 210093, China}
	\author{Yuqian Dong}
	\affiliation{National Laboratory of Solid State Microstructures, School of Physics,
		Nanjing University, Nanjing 210093, China}
	
	\author{Wen Zheng}
	\affiliation{National Laboratory of Solid State Microstructures, School of Physics,
		Nanjing University, Nanjing 210093, China}
	\author{Zhimin Wang}
	\affiliation{National Laboratory of Solid State Microstructures, School of Physics,
		Nanjing University, Nanjing 210093, China}
	\author{Xiangmin Yu}
	\affiliation{National Laboratory of Solid State Microstructures, School of Physics,
		Nanjing University, Nanjing 210093, China}
	\author{Dong Lan}
	\affiliation{National Laboratory of Solid State Microstructures, School of Physics,
		Nanjing University, Nanjing 210093, China}
	
	\author{Xinsheng Tan}
	\email{meisen0103@163.com}
	\affiliation{National Laboratory of Solid State Microstructures, School of Physics,
		Nanjing University, Nanjing 210093, China}
	
	\author{Yang Yu}
	\email{yuyang@nju.edu.cn}
	\affiliation{National Laboratory of Solid State Microstructures, School of Physics,
		Nanjing University, Nanjing 210093, China}

	\begin{abstract}	
		{Fast robust two-qubit gate operation with low susceptibility to crosstalk are the key to scalable quantum information processing. Parametrically driven gate is inherently insensitive to crosstalk while superadiabatic control can speed up the gate without losing accuracy. We propose and experimentally implement superadiabatic two-qubit gates using parametric modulation on superconducting quantum circuits. Our results demonstrate the preservation of adiabaticity at a gate speed close to the quantum limit, in addition to robustness against control instability. We demonstrate a CZ gate with error rate of 5.8$\%$, limited largely by qubit decoherence, promising future improvement and scalable implementation.}	
	\end{abstract}
	

	\maketitle

	High fidelity two-qubit quantum gates are a key component in quantum information processing~\cite{nielsen2002quantum,ladd2010quantum,Bennett1995Quantum}.
	The gate fidelity is damaged by both decoherence processes and control imperfections. While improving coherence is a long-term effort that requires upgrades in fabrication processes and extensive studies in materials, imperfect control due to crosstalk, instrumental instability, signal distortion, etc.\ is relatively remediable~\cite{strauch2003quantum,wendin2017quantum,krantz2019quantum}. In fact, recent development shows that complex and exquisite calibration processes are critical to battle these hurdles when controlling even a medium-scale quantum processor~\cite{arute2019quantum}. To alleviate such humongous efforts, a more robust gate scheme is desirable for further scaling up the system.
	
	Fast frequency modulation using flux pulse is widely used in superconducting qubit system~\cite{dicarlo2009demonstration,martinis2014fast}. Such control signal, if coupled to other qubits, will erroneously tune other qubits as well. In addition, pulse distortion adds transients to the actual signal, causing further complication and sometimes even confusion in the calibration step. Parametric gates relying on sideband driving circumvent this problem by activating qubit-qubit interactions with microwave pulses, and have been demonstrated to be a viable technique ~\cite{mckay2016universal,caldwell2018parametrically,didier2018analytical}. These gates also help alleviate the frequency crowding problem by making the interactions frequency-selective. Also, superadiabatic control technique is a powerful tool to expedite gate operations without damaging the accuracy and at no additional hardware cost~\cite{born1928beweis,berry1990histories,berry2009transitionless,an2016shortcuts,du2016experimental,wang2019experimental}. By taking advantage of both parametric gate and superadiabatic control, one expects to achieve better two-qubit gate performance.
	
	In this work, we propose a protocol to implement two-qubit superadiabatic (TQSA) gates with a parametric modulation scheme. In our scheme, a parametric modulating field provides fully tunable coupling between two qubits~\cite{zhou2009quantum,liu2014coexistence,wu2018efficient,caldwell2018parametrically,reagor2018demonstration,strand2013first,li2018perfect}, enabling us to construct a target superadiabatic Hamiltonian. We experimentally demonstrate TQSA gates in superconducting circuits consisting of multiple qubits. Using superadiabatic evolution we implement both SWAP gate and CZ gate. We track the state evolution in the \{$|01\rangle\,,\,|10\rangle$\} subspace and find no nonadiabatic error during the SWAP operation. Then we investigate the robustness of TQSA gates against the variations of control parameters. A superadiabatic CZ gate is finally demonstrated with a fidelity of 94.2$\%$, which is mainly limited by decoherence. Using numerical simulation, we prove that gate fidelity can reach 99.9$\%$, which is promising for quantum information processing.

	The principle of our protocol is as follows.
	We first introduce the target superadiabatic Hamiltonian. A two-level system coupled with a microwave field with frequency $\omega_m(t)$ and phase $\varphi(t)$ can be generally expressed as
	\begin{eqnarray}
	{H_0(t)=\frac{\hbar}{2}
		\begin{bmatrix}
		\varDelta(t) & \varOmega_{R}(t)e^{-i\varphi(t)} \\
		\varOmega_R (t)e^{i\varphi(t)} & -\varDelta(t)
		\end{bmatrix}},
	\label{eq:10}
	\end{eqnarray}
	where $\varDelta(t)=\omega_0-\omega_m(t)$ represents the detuning between energy gap of the two-level system and the frequency of the microwave field. $\varOmega_{R}(t)$ is the Rabi frequency, which is proportional to the amplitude. The instantaneous eigenvalues are
	$ E_{\pm}=\pm \frac{\hbar}{2} \sqrt{\varOmega_{R}(t)^2+\varDelta(t)^2}$. 
	For simplicity, we choose $\varphi(t)$ to be constant, and the auxiliary Hamiltonian $H_1$ in superadiabatic theory can be derived as
	\begin{eqnarray}
	{H_1(t)=\frac{i\hbar}{2}
		\begin{bmatrix}
		0 & {-\dot{\theta}(t)e^{-i\varphi}} \\
		{\dot{\theta}(t)e^{i\varphi}} & 0
		\end{bmatrix}}.
	\label{eq:21}
	\end{eqnarray}
	where $\theta(t)=\arctan(\varOmega_{R}(t)/\varDelta(t))$. Therefore, we obtain the Hamiltonian to implement a superadiabatic gate~\cite{liang2016proposal,zhang2017measuring}
	\begin{eqnarray}
	\begin{aligned}
	H_S(t)
	& =H_0(t)+H_1(t) \\
	& =\frac{\hbar}{2}
	\begin{bmatrix}
	\varDelta(t) & \varOmega_{S}(t)e^{-i(\varphi+\phi_S(t))} \\
	\varOmega_S (t)e^{i(\varphi+\phi_S(t))} & -\varDelta(t)
	\end{bmatrix},
	\label{eq:15}
	\end{aligned}
	\end{eqnarray}
	where $\varOmega_{S}(t)=\sqrt{\varOmega_{R}(t)^2+\dot{\theta}(t)^2}$, and $\phi_S(t)=\arctan[\dot{\theta}(t)/\varOmega_{R}(t)]$.

	We use parametric modulation to construct $H_S(t)$ in a subspace of a two-qubit system. Hamiltonian of two coupled qubits with one of them modulated by a longitudinal field $\varepsilon(t)$ can be written as
	\begin{eqnarray}
	{H=\frac{\hbar}{2}\sum\limits_{i=1}^{2}\omega_i\sigma_z^i+
		\hbar g(\sigma_+^1\sigma_-^2+\sigma_-^1\sigma_+^2)+
		\frac{\hbar}{2}f(\epsilon(t))\sigma_z^1},
	\label{eq:1}
	\end{eqnarray}
	where $\sigma_{z}^i$ is the Pauli operators and $\sigma_+^i$ ($\sigma_-^i$) is the creation (annihilation) operator in the Hilbert space of $i$th qubit $Q_i$. $g$ is the coupling strength between $Q_1$ and $Q_2$. $f(\varepsilon)$ is the nonlinear frequency response to the modulation pulse, and can be determined experimentally~\cite{supplementary1}. Here we choose $F(t)=\int{f(\varepsilon(t))dt}=A(t)\sin[(\omega_2-\omega_1)t+\delta_L(t)+\beta_L(t)]$ as an adjustable sinusoidal function intentionally, where $\omega_{1}$ ($\omega_2$) is the energy level spacing of $Q_1$ ($Q_2$), $\delta_L(t)$ and $\beta_L(t)$ are related to frequency detuning and phase of the longitudinal field, respectively. Applying unitary transformation~\cite{supplementary2}, we can rewrite Eq.~(4) as

	\begin{eqnarray}
	{H_I(t)=\frac{\hbar}{2}
		\begin{bmatrix}
		\ 0 & 0 & 0 & 0 \\
		\ 0 & \dot{\delta_{L}}(t) & 2gJ_{1}(A(t))e^{i\beta_L(t)} & 0 \\
		\ 0 & 2gJ_{1}(A(t))e^{-i\beta_L(t)} & -\dot{\delta_{L}}(t) & 0 \\
		\ 0 & 0 & 0 & 0
		\end{bmatrix}},
	\label{eq:5}
	\end{eqnarray}
	where $J_1$ is the first order Bessel function.

	Combining Eq.~(3) and Eq.~(5), we can calculate the parameters in $F(t)$ and the modulated pulse $\varepsilon(t)=f^{-1}(\dot{F}(t))$ to implement arbitrary TQSA gates in \{$|01\rangle$, $|10\rangle$\} subspaces. Similarly, we can construct Hamiltonian $H_0(t)$ using Eq.~(5) with specific $F(t)$ to realize two-qubit adiabatic (TQA) gates. It is worth emphasizing that if one considers the higher coupling energy level of transmons, such as $|11\rangle$ and $|20\rangle$, a similar Hamiltonian can be constructed to realize a CZ gate, as discussed later.
	
	\begin{figure}
		\begin{minipage}[b]{0.5\textwidth}
			\centering
			\includegraphics[width=8.5cm]{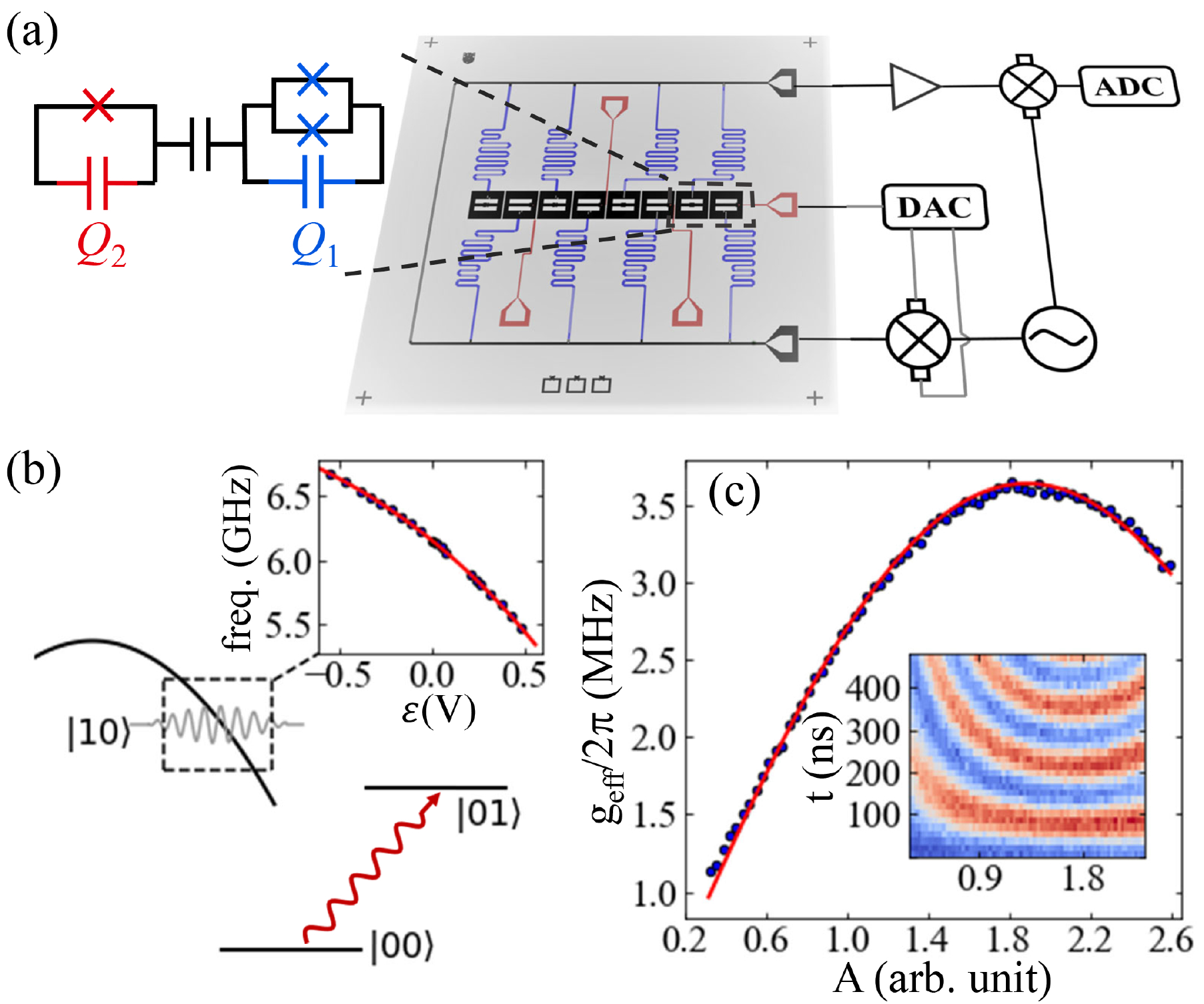}
		\end{minipage}
		\caption{(Color online) (a)$\,$False color image of the superconducting-qubit chip and equivalent circuit of the operating transmons. The schematic of control and measurement circuits is plotted on the right. The equivalent circuit of two operating transmons, labeled $Q_1$ and $Q_2$, is on the left. (b)$\,$Energy diagram for the TQSA operation. State $|10\rangle$ ($|01\rangle$) represents the first excited state of $Q_1$ ($Q_2$), while another qubit stays in its ground state. The frequency of state $|10\rangle$ is parametrically modulated as the function $f(\varepsilon(t))$ by specific microwave pulse applied through its Z control line. The upper right corner zooms in to the nonlinear frequency response to the flux pulse $f(\varepsilon)$ of $Q_1$. The red solid line shows the best fit with the cubic function. (c)$\,$Effective coupling strength controlled by parametric frequency modulation $f(t)=\frac{d}{dt}A\sin(\omega_1-\omega_2)t$. Insert shows the oscillation between $|10\rangle$ and $|01\rangle$ with different modulation amplitude $A$. $\rm g_{eff}$ extracted from the period of oscillations in insert are shown as blue dots. They fit well with first-order Bessel function $\rm g_{eff}=J_1(A)$ (red solid line). \label{fig:FIG1}}
	\end{figure}
	
	We demonstrate our protocol using superconducting quantum circuits~\cite{clarke2008superconducting,you2011atomic,gu2017microwave}. The chip contains eight transmons arranged in an array with nearest-neighbor coupling~\cite{koch2007charge}. Each qubit can be readout with an individual resonator which is coupled to the transmission line on the chip. For four of the eight transmons we replace the Josephson junction with two junctions in parallel (DC SQUID). Therefore, the frequency of those transmons can be tuned by applying pulses through the Z control lines, represented by red lines in Fig.~1(a).
	
	We demonstrate tunable coupling hence the TQSA quantum gate in two coupled qubits $Q_1$ and $Q_2$. $Q_2$ is a fixed frequency qubit with frequency $\omega_2/2\pi$ = 5.9498 GHz. The frequency of $Q_1$ can be tuned by combining static bias and fast flux pulse introduced through the Z control line [Fig.~1(a)]. In our protocol, energy spacing of qubit $Q_1$ is  statically set as $\omega_{1}/2\pi$ = 6.1567 GHz. In Fig.~1(b) and (c) we show parametric control of the effective coupling strength between $Q_1$ and $Q_2$.  The intrinsic coupling strength $g/2\pi$ between $Q_1$ and $Q_2$ is 6.26 MHz, determined by the capacity between the pads of transmons. We operate the system in the dispersive regime where $\Delta>>g_{c-q}$~, where $\Delta$ and  $g_{c-q}$ are detuning and coupling strength between cavity and qubit respectively \cite{blais2004cavity}. Both $Q_1$ and $Q_2$ are coupled to the readout transmission line which is also used for delivering the microwave signal for $XY$ control. The relaxation (dephasing) time of $Q_1$ is $T_1$ = $4.06$ $\mu$s ($T_\phi$ = 620 ns) at operation points. For $Q_2$, $T_1$ = 3.98 $\mu$s and $T_\phi$ = 6.1 $\mu$s. From Fig.~1(c) we find that the maximum effective coupling is about 3.6 MHz, leading to a quantum limit $\rm T_{QL}=\pi/2g_{eff}^{max}$ = 69 ns. Therefore, the minimum SWAP gate time for the dynamical scheme is about 70 ns.

	Having realized parametric tunable coupling, we can experimentally implement a specific TQSA gate and verify the acceleration of the superadiabatic scheme compared to the adiabatic process.
	We set the typical time dependent parameters of the TQSA gate as
	\begin{equation}
	\begin{aligned}
	& \varOmega_{R}(t)=\varOmega_{0}\sin(\frac{t}{T}\pi) \\
	& \varDelta(t)=\varOmega_{0}\cos(\frac{t}{T}\pi),
	\label{eq:9}
	\end{aligned}
	\end{equation}
	where $T$ is the gate duration time and $\varphi$(t) in Eq.~(1) is set as zero. In order to maximize efficiency we chose $T$ = 80 ns, which is subjected to the limitation of the maximum effective coupling strength. Using Eq.~(2), we obtain the auxiliary Hamiltonian $H_1=\varOmega_{A}(t)\sigma_{y}$, where $\varOmega_{A}(t)=\frac{\hbar\pi}{2T}$ is a constant value, as shown in Fig.~2(b). Using Eq.~(3) and Eq.~(5), we can calculate the specific form of $F(t)$ for our experiment with the parameters of $Q_1$ and $Q_2$~\cite{supplementary1}.
	
\begin{figure}
	\centering
	\begin{minipage}[b]{0.5\textwidth}
		\centering
		\includegraphics[width=8.5cm]{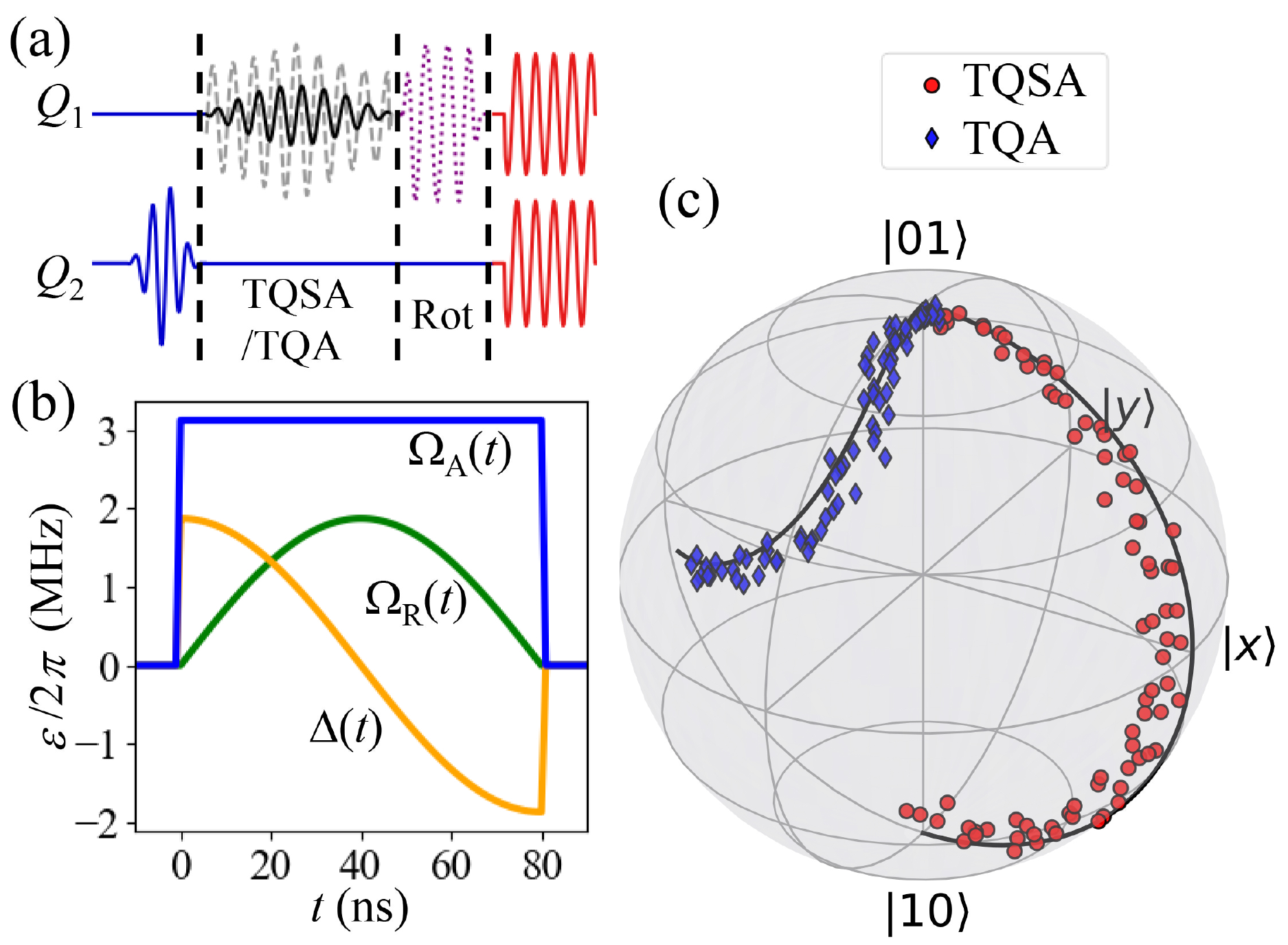}
	\end{minipage}
	\caption{(Color online) (a)$\,$Time profile for measuring the evolution trajectory of system states. Dashed vertical lines delineate the four steps of the experiment. The system is initialized in $|01\rangle$ with a $\pi$ pulse on $Q_2$ (blue). Then we implement the TQSA (TQA) gate, shown as light gray dashed (black solid) line. Subsequent $X/2,Y/2,I$ rotations, realized by dynamic parametric gates, project system states to three axes in $\{|01\rangle$, $|10\rangle\}$ subspace (purple dotted line). Finally we apply the readout pulse (red solid). (b)$\,$The typical component values of $H_S(t)$ in the parameter space to realize the TQSA protocol in experiments. $H_S(t)=\frac{\hbar}{2}(\varOmega_{R}(t)\sigma_{x}+\varOmega_{A}(t)\sigma_{y}+\varDelta(t)\sigma_{z})$. (c)$\,$Experimental results shown on the Bloch sphere represent the evolution of the quantum state in the subspace of \{$|01\rangle$, $|10\rangle$\}. Red dots (blue diamonds) represent experimental data of the superadiabatic (adiabatic) scheme, while black solid lines represent the numerical simulation of the two schemes.\label{fig:FIG2}}
\end{figure}

\begin{figure}
	\centering
	\includegraphics[width=8.5cm]{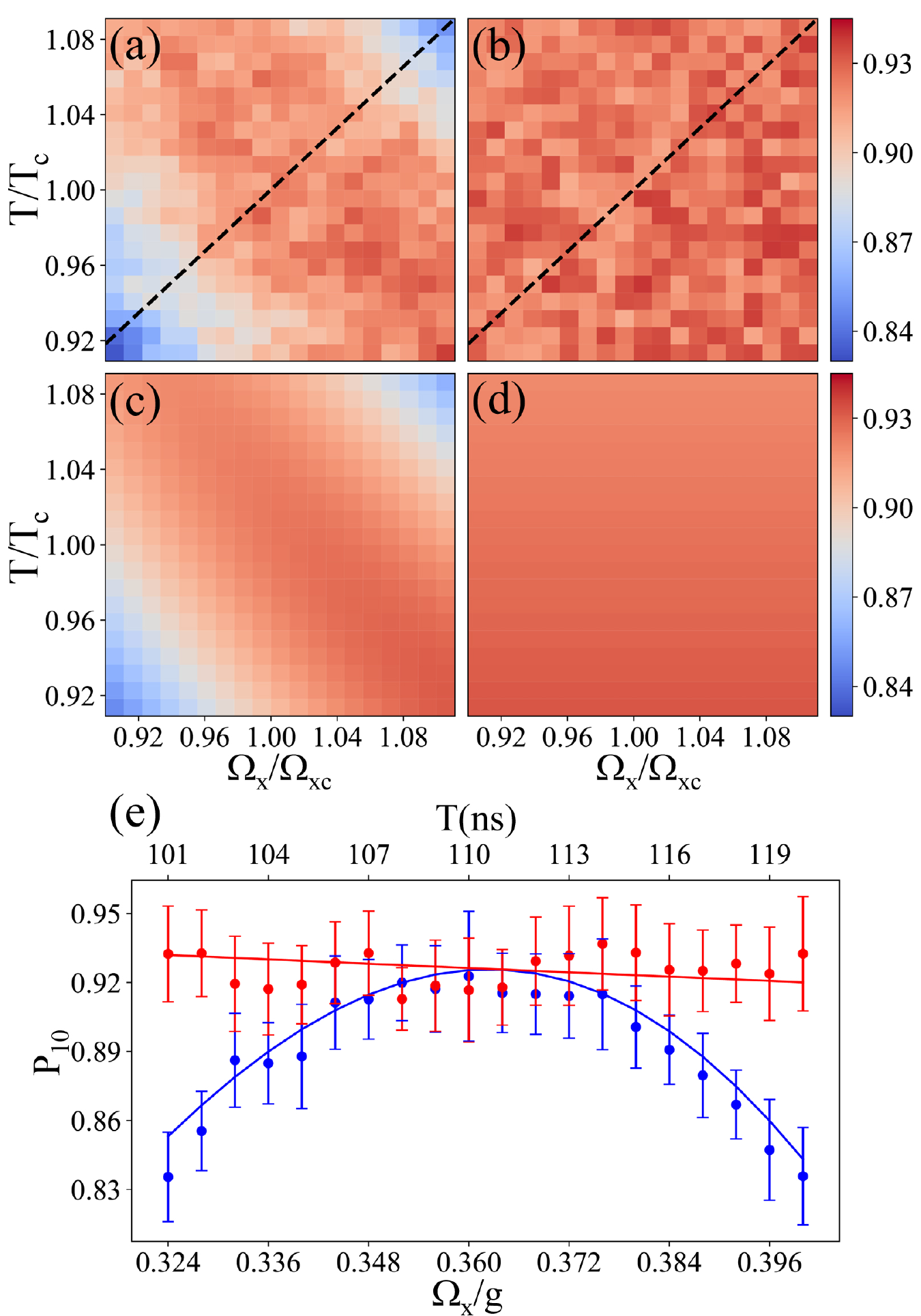}
	\caption{(Color online) Fidelity of two-qubit quantum SWAP gate as functions of control parameters in superadiabatic and dynamical schemes. (a)$\,$Contour plot of transfer fidelity, represented by the population of $|10\rangle$, as a function of normalized parameters $\varOmega_{x}/\varOmega_{xc}$ and $T/T_c$ for the dynamical scheme. (b)$\,$ Contour plot of transfer fidelity as a function of normalized parameters $\varOmega_{x}/\varOmega_{xc}$ and $T/T_c$ for the superadiabatic scheme. (c) and (d) are numerical simulation results involving the effect of decoherence. (e)$\,$ Cross section along the black dashed line in (a) and (b). Blue and red circles represent experimental fidelity of dynamical and superadiabatic SWAP gates respectively. Red solid (blue dashed) line is the numerical simulation considering decoherence.  \label{fig:FIG3}}
\end{figure}

	We track the system trajectory to verify the adiabaticity of the evolution. Time profile of the experiments is shown in Fig.~2(a). We apply a microwave pulse to $Q_2$, preparing the system in $|\psi(0)\rangle=|01\rangle$. TQSA(TQA) gate is then performed.  We use dynamic parametric gates to project system states to three axes in the subspace $\{|01\rangle$, $|10\rangle\}$. Finally we measure the system occupation probability of different states using two-qubit joint readout protocol~\cite{filipp2009two,dicarlo2009demonstration}. The result is shown in a Bloch sphere of $\{|01\rangle$, $|10\rangle\}$, with states out of the subspace is small and neglected~\cite{supplementary1}. By applying $H_S(t)$ to the qubits we realize superadiabatic operation. In Fig.~2(c) we show the state evolution trajectory of the TQSA gate in the Bloch sphere spanned by $|01\rangle$ and $|10\rangle$. The qubit state evolves precisely along the meridian predicted by the adiabatic theorem, proving the validity of the TQSA gate. It is noteworthy that the whole procedure time takes 80 ns, which is close to the quantum limit $T_{QL}$. The 10 ns extra time comes from the requirement of the protocol since we use a parametric pulse to control coupling strength. Furthermore, we compare our TQSA approach with the TQA routine with the same duration, as shown in Fig.~2(c). The TQA approach requires the fulfilling of adiabatic restriction, which is $T\geq10T_{QL}=690$ ns. Without auxiliary Hamiltonian $H_1$, the evolution trajectory deviates dramatically from the designed adiabatic path, which is caused by unwanted transitions between eigenstates of $H_0(t)$. The experimental results indicate that our TQSA scheme successfully accelerates the adiabatic procedure.
	
	\begin{figure}
		\centering
		\begin{minipage}[b]{0.5\textwidth}
			\includegraphics[width=8.5cm]{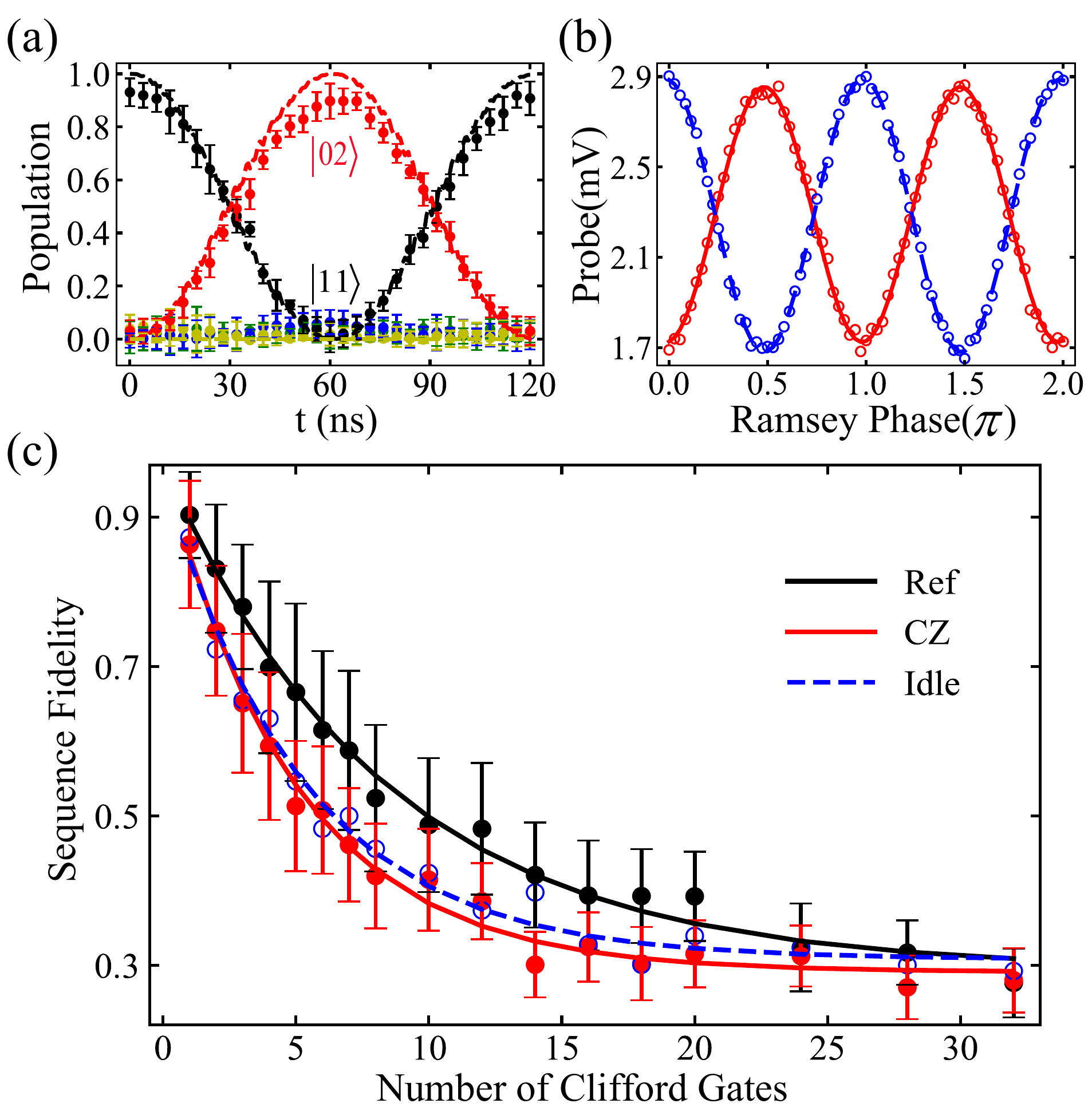}
		\end{minipage}
		\caption{(Color online) (a) States transfer during the SA-CZ gate. Red (black) dots represent population of $|02\rangle$ ($|11\rangle$), with standard derivation displayed as error bars. Populations of state $|00\rangle$ ($|01\rangle$, $|10\rangle$), arisen from imperfect state preparation and decay, are shown in green (blue, yellow) dots. Dashed lines are numerical simulation. (b) Ramsey type experiment with control qubit ($Q_1$) prepared in $|0\rangle$ (blue circles) or $|1\rangle$ (red circles) and $Q_2$ prepared in superposition state. A $\pi/2-pulse$ with changed phase is applied to $Q_2$ after the flux pulse. $Q_2$ is finally measured to determined the conditional phase acquired during the controlled-phase gate. Red-solid (blue-dashed) lines are corresponding sinusoidal fitting. (c) Interleaved randomized benchmarking result for SA-CZ gate. Reference (black) and interleaved (red) sequence fidelities are displayed as functions of number of Clifford gates. Decoherence error is quantified by idling for same duration as the SA-CZ gate, showed in blue dashed line. Each sequence fidelity is averaged over $k$ = 60 randomized operations.\label{fig:FIG4}}
	\end{figure}
	
	Compared to traditional two-qubit gates based on the dynamical procedure~\cite{caldwell2018parametrically}, TQSA gates possess the advantage of robustness against parameter fluctuations. The two important parameters for high fidelity gate operations are evolution time $T_c$ and Rabi frequency $\varOmega_{xc}$. Here $\varOmega_{xc}$ corresponds to the amplitude of the parametric field $F(t)$. In dynamic parametric scheme, the accuracy of gate operation is determined by both $T_c$ and $\varOmega_{xc}$. Therefore, the fluctuations of system parameters will significantly affect gate fidelity. To quantify the robustness of gate operation, we performed a SWAP gate using both superadiabatic and dynamical protocols. The artificial perturbations $\epsilon_{\varOmega_{x}}$ and $\epsilon_T$ are intentionally added. We choose $\epsilon_{\varOmega_{x}}\in[-0.1\varOmega_{xc},0.1\varOmega_{xc}]$ and $\epsilon_T\in[-0.1T_{c},0.1T_{c}]$.
	In experiments, we initialize the system state in $|01\rangle$, and set $\varOmega_{xc}$ = 0.36 g and $T_c$ = 110 ns. With varying $\epsilon_{\varOmega_{x}}$ and $\epsilon_T$, we measure the populations in state $|10\rangle$, which specify the gate fidelity. Fig.~3(a) and 3(b) show fidelity as functions of $\varOmega_{x}$ and $T$ with the superadiabatic and dynamical approaches respectively. Fidelity of the superadiabatic gate is more robust against the fluctuations of operation parameters, while the dynamical gate fidelity drops remarkably with increasing perturbations. In order to simultaneously display the influence of two parameters on fidelity, we show the cross section along the dashed line in Fig.~3(a) and 3(b). As expected, the 1D plot clearly indicates that TQSA gates are insensitive to control parameters compared to QDLG gates.

	To prove the generality of our protocol, we extend TQSA gate to \{$|11\rangle,|20\rangle$\} subspace, hence realize a superadiabatic-CZ (SA-CZ) gate. The SQUID $Q_1$ is biased at 6.4873 GHz with $\eta_{1}=-299.2$ MHz. The modulation frequency equals to {$\omega_{11}-\omega_{20}$}. The coupling strength between $|11\rangle$ and $|20\rangle$ is ${\sqrt{2}}g/2\pi$ = 9.14 MHz. We choose $T=60$ ns in Eq.~(6) (the limit time $T_{QL}$ is 47 ns). SA-CZ gate is realized by transferring $|11\rangle$ to $|20\rangle$ and back during a evolution time of 2$T$, as shown in Fig.~4(a). This procedure accumulates a conditional $\pi$-phase on state $|11\rangle$. A Ramsey type experiment is performed to verify the conditional phase, shown in Fig.~4(b). Fidelity of the SA-CZ gate is measured by interleaved random benchmarking (IRB)~\cite{barends2014superconducting,knill2008randomized,magesan2012efficient}. The occupation probability of state $|00\rangle$, which determines sequence fidelity, is measured as a function of the number of Clifford gates. We use fitting function $ P_{|00\rangle}(m)=Ap^m+B$ to extract the depolarizing parameter $ p_{ref}$ and $p_{CZ}$. The error rate is calculated with $ r_{SA-CZ}=(1-p_{CZ}/p_{ref})(d-1)/d$, where $d=4$ is the Hilbert-space dimension of two-qubit system. We obtain a $94.2\%$ SA-CZ gate fidelity. The decoherence contribution to gate error is measured by interleaving an idle of same duration as the CZ gate~\cite{barends2014superconducting}, as shown in Fig.~4(c). The idle fidelity is calculated to be $95.0\%$. We can tell that gate error is mainly caused by decoherence. Using numerical simulation, we prove that the SA-CZ gate fidelity can reach 99.94$\%$ in absence of decoherence~\cite{supplementary2}, limited by high order coupling in Jacobi-Anger expansion and small variation of coupling parameter g during frequency modulation of transmon qubits~\cite{didier2018analytical}.
	
	In summary, we propose and demonstrate TQSA gates using a parametric modulation protocol in superconducting circuits. The parametric gates can alleviate frequency crowding problems and circumvent the calibrations for pulse distortion and flux crosstalk. Using the parametric field, we modulate the coupling strength and phase to construct a superadiabatic Hamiltonian. The superadiabatic gate follows the expected adiabatic trajectory at a speed close to the quantum limit, exhibiting robustness against system or random fluctuations. The combined high fidelity and fast gate speed makes this TQSA gate promising for quantum information research.
	
	The authors thank H. Yu, Q. Liu and G. Xue for technical support. The authors also thank F. Yan for helpful discussion and improving the manuscript. This work was supported by the Key-Area Research and Development Program of GuangDong Province (Grant No. 2018B030326001), the NKRDP of China (Grant No. 2016YFA0301802), and the NSFC (Grants No.11604103, No. 11474153, and No. 91636218).

	\section{references}
	 

\end{document}